\documentclass[aps,prd,showpacs,twocolumn,bibnotes,nofootinbib]{revtex4-1}

\usepackage{amssymb}    
\usepackage{amsbsy}\newcommand{\mbf}{\boldsymbol}
\usepackage{amsmath}
\usepackage{amsfonts}
\usepackage{mathrsfs}
\usepackage{latexsym}
\usepackage[]{graphicx}
\usepackage[english]{babel}
\usepackage{url}
\usepackage{feyn}
\usepackage{MnSymbol}

\newcommand{\feynslash}[1]{#1\kern-0.45em/}

\begin{document}

\title{No one-loop gauge anomalies for a Lorentz-violating quantum electrodynamics: Evaluation of the three-photon vertex}

\author{D.H.T. Franco}
\email{daniel.franco@ufv.br (D.H.T. Franco)}
\affiliation{Universidade Federal de Vi\c{c}osa (UFV),
Departamento de F\'{i}sica, Campus Universit\'{a}rio,
Avenida Peter Henry Rolfs s/n - 36570-000 - Vi\c{c}osa - MG - Brazil.}

\author{A.H. Gomes}
\email{andre.herkenhoff@ufv.br (A.H. Gomes)}
\affiliation{Universidade Federal de Vi\c{c}osa (UFV),
Departamento de F\'{i}sica, Campus Universit\'{a}rio,
Avenida Peter Henry Rolfs s/n - 36570-000 - Vi\c{c}osa - MG - Brazil.}

\date{\today}

%
\begin{abstract}
Identification of the diagrams that can lead to gauge anomalies in the (minimal) Lorentz- and $CPT$-violating extension of quantum electrodynamics reveals these are the electron self-energy and vertex correction (related to the Ward-Takahashi identity), the photon self-energy (related to the vacuum polarization tensor transversality), and the three-photon vertex diagrams. All but the latter were explicitly verified to be free of anomalies to first order in loop expansion. Here we provide this remaining evaluation and verify the absence of anomalies in this process.
\end{abstract}

\maketitle

\section{Introduction}
\label{introduction}

Planck scale phenomena are expected to be described by fundamentally new physics other than the traditional one describing our relatively low-energy world. Due to the high energy involved, a long sought quantum theory of gravity may play fundamental role, along with the other known interactions, in describing this extreme environment, and spacetime local structure may reveal itself as behaving very differently from Minkowskian one. As the Universe expanded and cooled to its present state, high energy relic effects could reveal themselves as small deviations from known well established physics and, of these, Lorentz symmetry violation is a possibility that has been intensively studied in the past decades, leading to a vast spectrum of new theoretical constructions \cite{annual-review}.

A research program for Lorentz symmetry violations, initiated by Colladay and Kosteleck\'{y} \cite{sme1,sme2}, deals with a (quantum field theory based) standard model extension (SME) allowing for Lorentz and $CPT$ violations coming from all possible observer Lorentz scalars formed by couplings between standard model fields and coefficients with Lorentz indices, playing the role of constant background fields. Such a construction could arise, for instance, as an effective low-energy limit of a high-energy Lorentz invariant theory which undergoes spontaneous Lorentz symmetry breaking. The minimal extension exhibits the standard model SU(3)$\times$SU(2)$\times$U(1) gauge structure and is power-counting renormalizable, yielding a Lagrangian with a finite number of Lorentz noninvariant terms of mass dimension equal or less than four composed with standard model fields --- causality, unitarity, and other issues may be further analysed \cite{causality-unitarity,adam-klinkhamer,lehnert,k-f-1,k-f-2}.

Several interesting effects coming from Lorentz violation have been studied. Some of these deals with spacetime-varying couplings \cite{varying-couplings1,varying-couplings2}, alternative models for neutrino oscillations \cite{neutrino}, black hole thermodynamics \cite{black-hole1,black-hole2}, Lorentz violation as an origin for gravitation \cite{gravitation}, spontaneous breaking of Lorentz invariance leading to a nonlinear electrodynamics \cite{nonlinearelectrodynamics}, among others \cite{kost-proceedings}. On the other hand, high precision measurements have already placed stringent upper bounds on coefficients for Lorentz violation from various sectors \cite{data-table}, but with no claims for discovery of Lorentz violation so far. In a more formal aspect, consistency under quantum corrections of sectors of the SME has also been studied, for instance in Refs. \cite{kost-1-loop,ufjf,colladay-mcdonald,ferrero-altschul,algebraic-renorm}, and is the topic of the present work. 

Here, we give a natural continuation of a previous work \cite{algebraic-renorm}, where we investigated the renormalizability to all orders of the Lorentz noninvariant (minimal) extension of the quantum electrodynamics (QED) in the algebraic approach \cite{piguet-rouet,piguet}. There, we identified new anomaly structures besides the usual Adler-Bardeen-Bell-Jackiw (ABBJ) \cite{abbj-adler,abbj-bardeen,abbj-bell-jackiw} one. Now, we consider the issue of identifying from which processes the potential anomalies come from. Of these, the standard Ward-Takahashi identity relating the vertex correction and electron self-energy diagrams, and the transversality of the vacuum polarization are known to be satisfied to one-loop order \cite{kost-1-loop}. Here, we make further progress by evaluating the one-loop three-photon vertex diagram. We find it to respect gauge symmetry, assuring the absence of gauge anomalies to this order in perturbation theory.

This paper is presented as follows. In Section \ref{framework}, we introduce the QED extension allowing for Lorentz violation along with considerations for one-loop evaluations. In Section \ref{section:anomalies} we evaluate the one-loop three-photon vertex and verify the absence of (vectorial) gauge anomaly. Contributions for this process from multiloop and nonlinearity in Lorentz-violating coefficients are also briefly discussed. Our results are summarized in Section \ref{section:conclusions} and some perspectives are given.

\section{Framework}
\label{framework}

\subsection{QED extension}

The action $\mathcal{S}$ of the Lorentz- and $CPT$-violating extension of the QED for electrons, positrons, and photons is given by \cite{sme1}
\begin{align}\label{qedex}
\mathcal{S} = \int d^{\,4}x\, \Big[	& i\, \overline{\psi} (\gamma^\mu+\Gamma_1^\mu) D_\mu\, \psi - \overline{\psi}(m+M_1)\psi - \frac{1}{4}F^{\mu\nu} F_{\mu\nu}	\nonumber \\
& - \frac{1}{4}(k_F)_{\mu\nu\rho\sigma} F^{\mu\nu}F^{\rho\sigma} + (k_{AF})_\mu A_\nu \widetilde{F}^{\mu\nu}	\Big],
\end{align}
which, besides the usual QED  terms, includes Lorentz-breaking terms whose coefficients have the form of constant background fields --- see the last two terms and the following definitions:
\begin{align}
&	\Gamma_1^\mu \equiv c^{\,\lambda\mu}\gamma_\lambda + d^{\lambda\mu}\gamma_5\gamma_\lambda + e^{\,\mu} + 
if^\mu\gamma_5 + \frac{1}{2}\,g^{\kappa\lambda\mu}\sigma_{\kappa\lambda},	\nonumber	\\
&	M_1 \equiv im_5\gamma_5 + a_{\mu}\gamma^\mu + b_{\mu}\gamma_5\gamma^\mu + \frac{1}{2}\,H_{\mu\nu}\sigma^{\mu\nu}.
\end{align}
Terms with coefficients of even (odd) number of indexes respect (do not respect) $CPT$ symmetry. Coefficients $(k_F)_{\mu\nu\rho\sigma}$ and those appearing in $\Gamma_1^\mu$ are dimensionless, while $(k_{AF})_\mu$ and the ones in $M_1$ have dimension of mass. The trace part of $c_{\mu\nu}$ is Lorentz invariant and only yields a redefinition of the fermion fields, so we take it to be zero; $H_{\mu\nu}$ and the first two indices of $g_{\mu\nu\rho}$ are antisymmetric; $(k_F)_{\mu\nu\rho\sigma}$ have the symmetries of the Riemann tensor and, analogous to $c_{\mu\nu}$, can be taken double traceless; at last, $m_5$ can be eliminated by a chiral rotation in the absence of chiral anomalies. Along with action (\ref{qedex}), we may add
\begin{equation}\label{gf-ir}
	\mathcal{S}_{\mathrm{GF+IR}} = \int d^{\,4}x\, \left[	-\frac{1}{2\alpha}\left(\partial_\mu A^\mu\right)^2	+ \frac{1}{2}\mu^2 A_\mu A^\mu	\right],
\end{equation}
\textit{i.e.}, a gauge-fixing term and an infrared (IR) regulator, introduced in order to avoid infrared singularities by means of a mass term for the photon field, respectively.

We define a (local) gauge Ward operator,
\begin{equation}\label{localward}
w_{\rm g}(x) = -\partial^{\,\mu} \frac{\delta}{\delta A^\mu} + ie\left(	\frac{\overleftarrow{\delta}}{\delta\psi}\,\psi - \overline{\psi}\,\frac{\overrightarrow{\delta}}{\delta\overline{\psi}}	\right),
\end{equation}
to functionally implement gauge transformations. This is a symmetry of the action (\ref{qedex}), $w_{\rm g}(x) \mathcal{S} = 0$, and the addition of the gauge-fixing and IR regulator (\ref{gf-ir}) linearly breaks it,
\begin{equation}\label{classical-gauge-linear-breaking}
w_{\rm g}(x) \mathcal{S} = - \left( \frac{\Box+\alpha\mu^2}{\alpha} \right)\partial_\mu A^\mu(x),
\end{equation}
but due to this linearity, the right-hand side of this expression receives no quantum corrections during renormalization procedure, remaining a classical breaking.

\subsection{Setup for one-loop evaluations}
\label{section:setup}

Precision experiments place very stringent upper bounds on coefficients for Lorentz violation from various sectors of the SME \cite{data-table}, so we define as concordant frames those which move nonrelativistically with respect to the Earth, where these coefficients are measured to be very small. We restrict our analyses to these frames to avoid spurious enlargements of the coefficients and construct a meaningful perturbative expansion. Calculations will be done to first order in loop-expansion and only contributions linear in Lorentz-violating coefficients may be considered --- nonlinear contributions may be of the same order of magnitude as those from diagrams with higher number of loops, which we do not consider.

An important consequence of the smallness of possible Lorentz noninvariance effects is that we may regard the Lorentz-violating pieces of (\ref{qedex}) as interaction vertice. Therefore, we have the standard QED Feynman rules along with new ones for the Lorentz noninvariant vertices with coefficients for Lorentz violation entering as propagator or vertex insertions. Propagator insertions read:
\begin{equation}
\feyn{f a f x f a f} = - i M_1,
\end{equation}
\begin{equation}
\feyn{f a f f a f} \hspace{-1.55cm}\bullet \hspace{+1.4cm}= i \Gamma^\mu_1 p_\mu,
\end{equation}
\begin{equation}
\mu\,\,\, \feyn{ g g } \hspace{-.85cm}\bullet \qquad\nu \hspace{+.0cm}= - 2 i p^\alpha p^\beta (k_F)_{\alpha\mu\beta\nu},
\end{equation}
\begin{equation}
\mu\,\,\, \feyn{ g g } \hspace{-.8cm}\mbf{\bigtimes} \quad\,\,\,\,\,\nu \hspace{+.0cm} = 2 (k_{AF})^\alpha \varepsilon_{\alpha\mu\beta\nu} p^\beta,
\end{equation}
and the extra interaction vertex is given by:
\begin{equation}
\Diagram{ \\ fdV \\ & g \\ fuA \\ } \hspace{-.82cm}\bullet \hspace{+.7cm}= - i e \Gamma^\mu_1.
\end{equation}
%

\section{Searching for anomalies}
\label{section:anomalies}

In this section, we first identify from which diagrams gauge anomalies could emerge for a model based on (\ref{qedex}). These are the vertex correction and electron self-energy, photon self-energy, and three-photon vertex diagrams. The only one remaining to be explicitly studied in the literature is the three-photon vertex, which we deal with and verify it to be anomaly-free.

\subsection{Possible anomalies in the QED extension}

From the vertex functional $\Gamma$, defined as the generating functional of 1-particle irreducible graphs, we can read from it all potential anomalies of a model based on (\ref{qedex}). The one-loop quantum extension of the classical expression (\ref{classical-gauge-linear-breaking}) reads \cite{algebraic-renorm}
\begin{align}\label{ward-anomalies}
w_{\rm g}(x)\, \Gamma  = &- \left( \frac{\Box+\alpha\mu^2}{\alpha} \right)\partial_\mu A^\mu + \lambda^{(1)}\overline{\psi}\psi +\, 																						i\lambda^{(2)}\overline{\psi}\gamma_5\psi	\nonumber	\\
&				+\, \lambda^{(3)}_\mu\overline{\psi}\gamma^\mu\psi +\, \lambda^{(4)}_\mu\overline{\psi}\gamma^\mu\gamma_5\psi +\, 								\lambda^{(5)}_{\mu\nu}\overline{\psi}\,\sigma^{\mu\nu}\psi	\nonumber	\\
&				+\, \lambda^{(6)}_{\mu\nu}F^{\mu\nu} +\, \lambda^{(7)}_{\mu\nu\rho\sigma} F^{\mu\nu}F^{\rho\sigma} +\, \mathcal{O}\left(\hbar^2\lambda\right).
\end{align}
The first term at the right-hand side of (\ref{ward-anomalies}) is a linear breaking previously discussed and receives no quantum corrections. The others represent potential anomalies, with associated $\hbar$-order anomaly coefficients $\lambda^{(i)}$ as functions of parameters appearing in (\ref{qedex}). All potentially anomalous structures appearing come from gauge Ward identities of standard QED but generalized to consider possible violations, generally due to new tensor structures coming from Lorentz noninvariance and the lack of discrete symmetries%
\footnote{For comparison reasons, consider the analogous of (\ref{ward-anomalies}) for the standard QED, $w_{\rm g}\Gamma_{\rm QED}+\frac{1}{\alpha}\left( \Box+\alpha\mu^2 \right)\partial_\mu A^\mu=0$. The right-hand side of this identity is constrained to be zero, for instance, due to discrete $C$- and $PT$-symmetries respected by $\Gamma_{\rm QED}$ and Lorentz invariance, both restricting all appearing terms to have the form of a gauge variation of field polynomials, which are then further reabsorbed by $w_{\rm g}\Gamma_{\rm QED}$, and the polynomials interpreted as noninvariant counterterms coming from noninvariant regularization schemes.}%
.

It can be easily verified that evaluation of coefficients $\lambda^{(1)}$ to $\lambda^{(5)}$ hinges in the computation of the vertex correction $\Gamma_\mu(p,q)$ and electron self-energy diagrams $\Sigma(p)$,
\begin{align}\label{ward-takahashi}
& -q^\mu \Gamma_\mu(p,q) - e\Sigma(p+q) + e\Sigma(p)	\nonumber\\
&\quad = \lambda^{(1)} + i\lambda^{(2)}\gamma_5 + \lambda^{(3)}_\alpha\gamma^\alpha + \lambda^{(4)}_\alpha\gamma^\alpha\gamma_5 + \lambda^{(5)}_{\alpha\beta}\sigma^{\alpha\beta},
\end{align}
and that coefficient $\lambda^{(6)}$ is related to the photon self-energy $\Pi_{\mu\nu}(k)$,
\begin{equation}\label{vacuum-coeff}
\lambda_{\mu\nu}^{(6)}k^\mu = -\frac{1}{2}k^\mu\Pi_{\mu\nu}(k).
\end{equation}
These six coefficients vanish to one-loop order \cite{kost-1-loop}. Coefficient $\lambda^{(7)}$ is the only one remaining in the literature to be evaluated, and is related to the three-photon vertex, as we discuss next.

\subsection{One-loop three-photon vertex}

%
\begin{figure}[ht]
\begin{center}
\includegraphics[scale=.31]{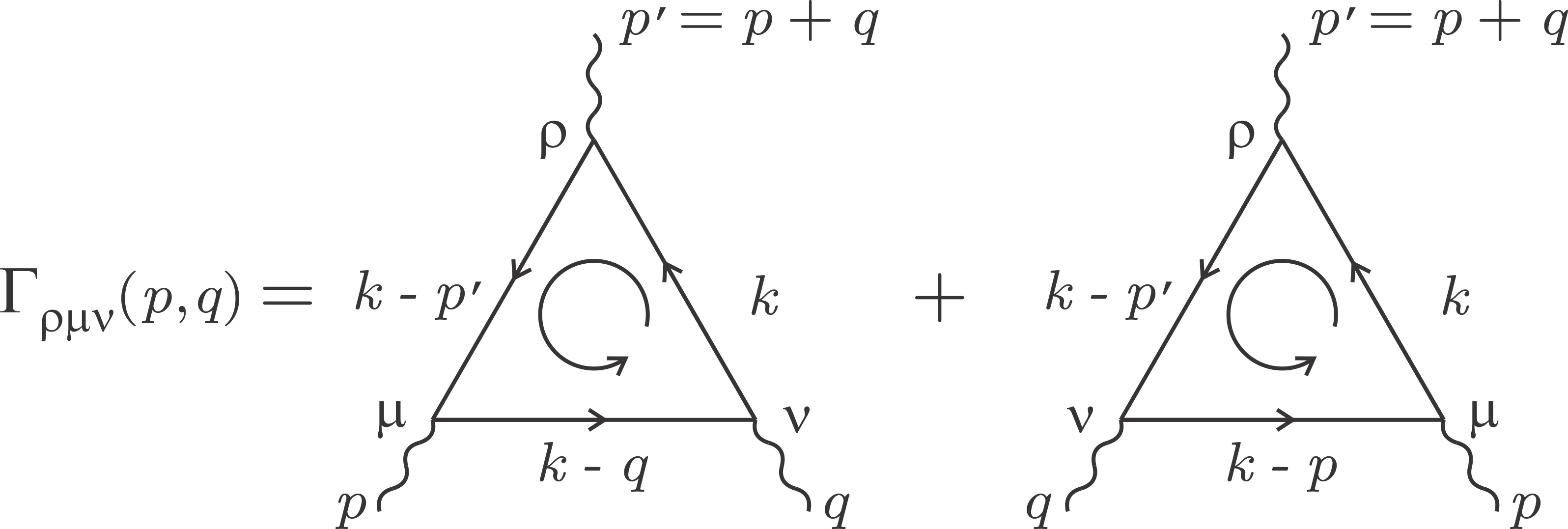}
\caption{One-loop three-photon vertex.}
\label{fig:triangular-two}
\end{center}
\end{figure}

From expression (\ref{ward-anomalies}), the following is obtained:
\begin{equation}\label{triang-first-eq}
(p+q)^\rho \Gamma_{\rho\mu\nu}(p,q) = -8i \lambda^{(7)}_{\rho\mu\sigma\nu} p^\rho q^\sigma,
\end{equation}
with external momenta as depicted in Fig. \ref{fig:triangular-two}. This expression relates $\lambda^{(7)}_{\rho\mu\sigma\nu}$ to the one-loop photon three-point function $\Gamma_{\rho\mu\nu}(p,q)$, and represents gauge current conservation if $\lambda^{(7)}_{\rho\mu\sigma\nu} = 0$, otherwise it is not conserved due to the nonvanishing anomaly. Instead of working on an expression for $\lambda^{(7)}_{\rho\mu\sigma\nu}$, in what follows we evaluate the left-hand side of (\ref{triang-first-eq}). Before, it is important to notice that in (\ref{ward-anomalies}), due to the contraction with $F^{\mu\nu}F^{\rho\sigma}$, only pieces of this coefficient respecting
\begin{equation}
\lambda^{(7)}_{\mu\nu\rho\sigma} = -\lambda^{(7)}_{\nu\mu\rho\sigma} = -\lambda^{(7)}_{\mu\nu\sigma\rho} = \lambda^{(7)}_{\rho\sigma\mu\nu}
\end{equation}
can contribute to the anomaly. Therefore, any piece without this index symmetries appearing after the evaluation of the left-hand side of (\ref{triang-first-eq}) will not represent an anomaly: it represents noninvariant counterterms that can be reabsorbed by a redefinition of the vertex functional $\Gamma$ at that loop order and are further cancelled, order by order in perturbation theory, leaving no physically measurable effect \cite{nair}. 

Bose symmetry must be taken into account and, as seen in Fig. \ref{fig:triangular-two}, we consider both diagrams along with a convenient internal momenta routing. According to generalized Furry's theorem \cite{kost-1-loop}, only C-odd insertions give nonzero contribution to processes with odd number of external photon legs attached to a fermion loop. Therefore, the nonvanishing part of $\Gamma_{\rho\mu\nu}(p,q)$ is given by the sum of all possible processes with one propagator or vertex $C$-odd insertion --- these are depicted in Fig. \ref{fig:triangular-insertions}. Contraction of the integral for $\Gamma_{\rho\mu\nu}(p,q)$ with $(p+q)^\rho$ leads to:
\begin{figure}[ht]
\begin{center}
\includegraphics[scale=.45]{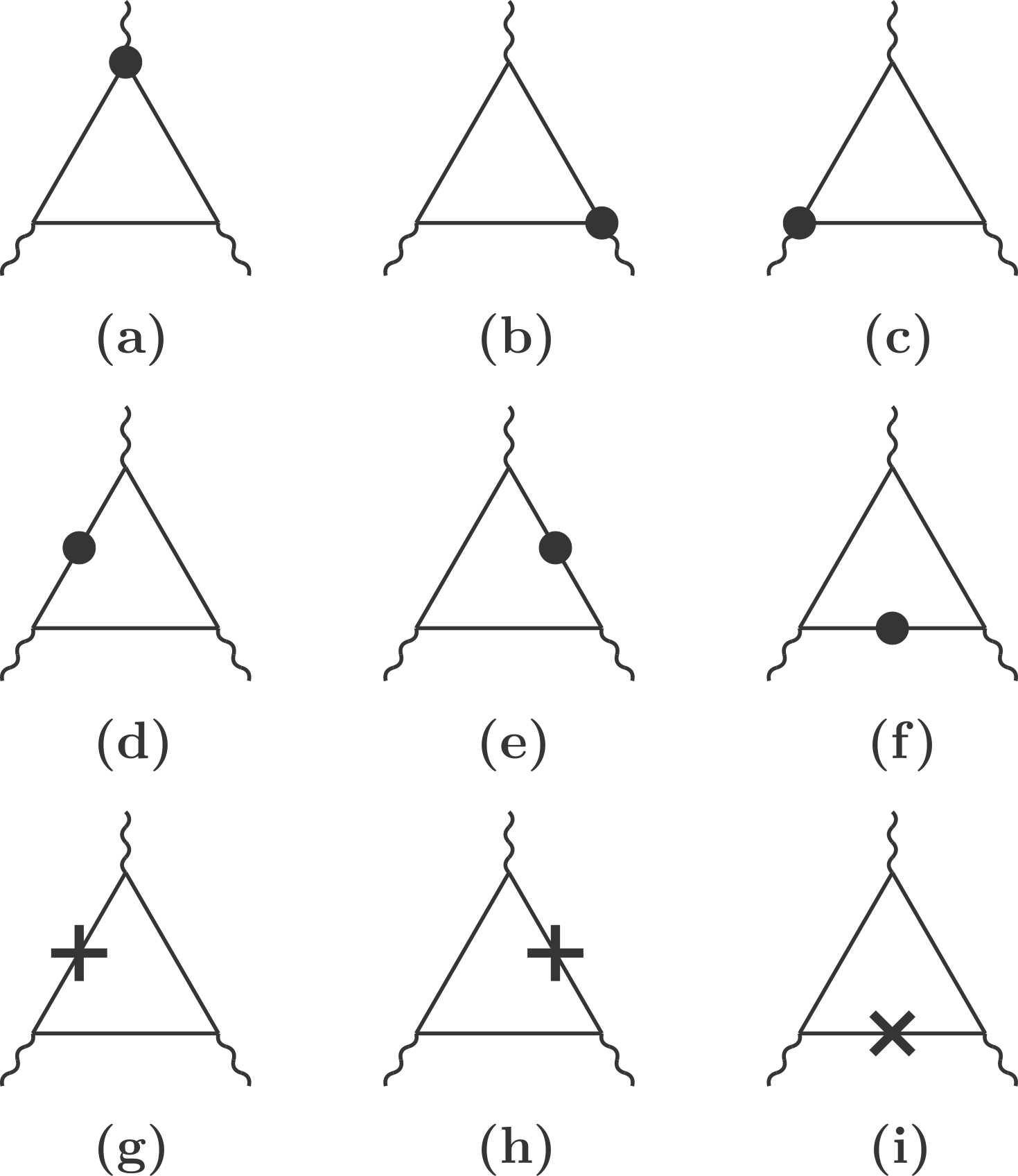}
\caption{Lorentz-violating insertions in the one-loop three-photon vertex. Generalization of Furry's theorem states that only $C$-odd insertions can give nonvanishing contributions to this process.}
\label{fig:triangular-insertions}
\end{center}
\end{figure}

\begin{widetext}
\begin{align}\label{triang-all-diagr}
(p+q)^\rho \Gamma_{\rho\mu\nu}(p,q) & = (p+q)^\rho \left(\Gamma_{\rho\mu\nu}^{(a)}+\cdots+\Gamma_{\rho\mu\nu}^{(i)}\right) \nonumber\\
	\displaybreak[0]
& = e^3 \int \frac{d^4k}{(2\pi)^4} \textrm{Tr}\Bigg\{
\Bigg[ \Gamma_{1\rho}p'^\rho\frac{1}{\feynslash{k}-\feynslash{p}'-m}\gamma_\mu\frac{1}{\feynslash{k}-\feynslash{q}-m}\gamma_\nu\frac{1}{\feynslash{k}-m} + (\mu;p\leftrightarrow\nu;q) \Bigg] \nonumber\\
	\displaybreak[0]
&\qquad\qquad\qquad\qquad + \Bigg[ \Gamma_{1\nu}\frac{1}{\feynslash{k}-\feynslash{p}'-m}\gamma_\mu\frac{1}{\feynslash{k}-\feynslash{q}-m} - \Gamma_{1\nu}\frac{1}{\feynslash{k}-m}\gamma_\mu\frac{1}{\feynslash{k}-\feynslash{q}-m} + (\mu;p\leftrightarrow\nu;q)  \Bigg] \nonumber\\
	\displaybreak[0]
&\qquad\qquad\qquad\qquad + \Bigg[ \Gamma_{1\nu}\frac{1}{\feynslash{k}-\feynslash{p}-m}\gamma_\mu\frac{1}{\feynslash{k}-\feynslash{p}'-m} - \Gamma_{1\nu}\frac{1}{\feynslash{k}-\feynslash{p}-m}\gamma_\mu\frac{1}{\feynslash{k}-m} + (\mu;p\leftrightarrow\nu;q)  \Bigg] \nonumber\\
	\displaybreak[0]
&\qquad\qquad\qquad\qquad + \Bigg[ -\Gamma_{1\lambda}(k-p')^\lambda\frac{1}{\feynslash{k}-\feynslash{p}'-m}\gamma_\mu\frac{1}{\feynslash{k}-\feynslash{q}-m}\gamma_\nu\frac{1}{k-\feynslash{p}'-m} \nonumber\\
	\displaybreak[0]
&\qquad\qquad\qquad\qquad\qquad + \Gamma_{1\lambda}(k-p')^\lambda\frac{1}{\feynslash{k}-\feynslash{p}'-m}\gamma_\nu\frac{1}{\feynslash{k}-\feynslash{p}-m}\gamma_\mu\frac{1}{\feynslash{k}-m} + (\mu;p\leftrightarrow\nu;q)  \Bigg] \nonumber\\
	\displaybreak[0]
&\qquad\qquad\qquad\qquad + \Bigg[ -\Gamma_{1\lambda}k^\lambda\frac{1}{\feynslash{k}-\feynslash{p}'-m}\gamma_\mu\frac{1}{\feynslash{k}-\feynslash{q}-m}\gamma_\nu\frac{1}{\feynslash{k}-m} \nonumber\\
	\displaybreak[0]
&\qquad\qquad\qquad\qquad\qquad + \Gamma_{1\lambda}k^\lambda\frac{1}{\feynslash{k}-m}\gamma_\nu\frac{1}{\feynslash{k}-\feynslash{p}-m}\gamma_\mu\frac{1}{\feynslash{k}-m} + (\mu;p\leftrightarrow\nu;q)  \Bigg] \nonumber\\
	\displaybreak[0]
&\qquad\qquad\qquad\qquad + \Bigg[ -\Gamma_{1\lambda}(k-q)^\lambda\frac{1}{\feynslash{k}-\feynslash{q}-m}\gamma_\nu\frac{1}{\feynslash{k}-\feynslash{p}'-m}\gamma_\mu\frac{1}{\feynslash{k}-\feynslash{q}-m} \nonumber\\
	\displaybreak[0]
&\qquad\qquad\qquad\qquad\qquad + \Gamma_{1\lambda}(k-p)^\lambda\frac{1}{\feynslash{k}-\feynslash{p}-m}\gamma_\mu\frac{1}{\feynslash{k}-m}\gamma_\nu\frac{1}{\feynslash{k}-\feynslash{p}-m} + (\mu;p\leftrightarrow\nu;q)  \Bigg] \Bigg\},
\end{align}
\end{widetext}
%
where the terms inside the first square brackets at the right-hand side are the contributions coming from $\Gamma_{\rho\mu\nu}^{(a)}$, the ones inside the second are the contributions from $\Gamma_{\rho\mu\nu}^{(b)}$ and so on until the last square brackets, which comprises the contributions from $\Gamma_{\rho\mu\nu}^{(f)}$. Diagrams (g), (h), and (i), involving propagator insertions of $M_1$, give vanishing contributions as will be explained below. First, we note the following:
\begin{itemize}
\item Shifting $k\rightarrow k-p$ in the second term from $\Gamma_{\rho\mu\nu}^{(b)}$, it cancels the first term from $\Gamma_{\rho\mu\nu}^{(c)}$;
\item Shifting $k\rightarrow k-q$ in the second term from $\Gamma_{\rho\mu\nu}^{(c)}$, it cancels the first term from $\Gamma_{\rho\mu\nu}^{(b)}$;
\item Shifting $k\rightarrow k-q$ in the second term from $\Gamma_{\rho\mu\nu}^{(e)}$, it cancels the first term from $\Gamma_{\rho\mu\nu}^{(f)}$;
\item Shifting $k\rightarrow k-q$ in the second term from $\Gamma_{\rho\mu\nu}^{(f)}$, it cancels the first term from $\Gamma_{\rho\mu\nu}^{(d)}$;
\item We have analogous results for the above with $\mu \leftrightarrow \nu$ and $p \leftrightarrow q$;
\item All the remaining nonmentioned terms add up to zero.
\end{itemize}
These are linearly divergent integrals so that shifts in the integration momenta generate relevant (finite) nonvanishing surface terms \cite{surface}:
%
\begin{widetext}
\begin{align}\label{triang-surface}
&(p+q)^\rho \Gamma_{\rho\mu\nu}(p,q) = e^3 \int \frac{d^4k}{(2\pi)^4} \textrm{Tr}\Bigg\{p^\alpha\frac{\partial}{\partial k^\alpha}\Bigg( \Gamma_{1\nu}\frac{1}{\feynslash{k}-m}\gamma_\mu\frac{1}{\feynslash{k}-\feynslash{q}-m}\Bigg) + q^\alpha\frac{\partial}{\partial k^\alpha}\Bigg( \Gamma_{1\nu}\frac{1}{\feynslash{k}-\feynslash{p}-m}\gamma_\mu\frac{1}{\feynslash{k}-m}\Bigg)	\nonumber\\
	\displaybreak[0]
& - q^\alpha\frac{\partial}{\partial k^\alpha}\Bigg( \Gamma_{1\lambda}k^\lambda\frac{1}{\feynslash{k}-m}\gamma_\nu\frac{1}{\feynslash{k}-\feynslash{p}-m}\gamma_\mu\frac{1}{\feynslash{k}-m} - \Gamma_{1\lambda}(k-p)^\lambda\frac{1}{\feynslash{k}-\feynslash{p}-m}\gamma_\mu\frac{1}{\feynslash{k}-m}\gamma_\nu\frac{1}{\feynslash{k}-\feynslash{p}-m}\Bigg) + (\mu;p\leftrightarrow\nu;q) \Bigg\}.
\end{align}
\end{widetext}
%
As stated before, propagator insertions of $M_1$ give vanishing contributions since their associated integrals are logarithmically divergent and the resulting surface integrals converge to zero when we take the surface at infinity. Also, only the $d^{\mu\nu}\gamma_5\gamma_\mu$ part of $\Gamma^\nu_1$ contribute%
\footnote{As mentioned earlier, only $C$-odd insertions are relevant. Of these, the $e_\mu$ part gives vanishing contribution because the trace of an odd number of $\gamma$ matrices vanishes and $f_\mu$ does not contribute because the trace of the product of $\gamma_5$ with an odd number of $\gamma$'s vanishes too, remaining only $d_{\mu\nu}$.}%
.

To evaluate integral (\ref{triang-surface}) we use Gauss' theorem to change it into a surface integral, which is easy to solve considering an isotropic (hyper-)surface at $k\rightarrow\infty$. Before, however, we need to evaluate traces involving the product of $\gamma_5$ with four and six $\gamma$'s. For the former we use $\textrm{Tr}(\gamma_5\gamma_\mu\gamma_\nu\gamma_\rho\gamma_\sigma) = -4i\varepsilon_{\mu\nu\rho\sigma}$ and for the latter we make extensive use of identities $\feynslash{a}\gamma_\mu\feynslash{b}=2\feynslash{a}b_\mu-\feynslash{a}\feynslash{b}\gamma_\mu$ and $\feynslash{a}\feynslash{b}=2a{\cdot}b-\feynslash{b}\feynslash{a}$ to reduce the trace to the first case. After evaluation of the traces and reorganizing, we find:
%
\begin{widetext}
\begin{align}\label{before-gauss}
(p+q)^\rho \Gamma_{\rho\mu\nu}(p,q) = 4ie^3 \int \frac{d^4k}{(2\pi)^4}\Bigg\{ & d^\lambda_{\,\,\,\nu}\varepsilon_{\lambda\sigma\mu\rho}(p^\sigma q^\alpha - p^\alpha q^\sigma)\frac{\partial}{\partial k^\alpha}\Bigg[\frac{k^\rho}{(k^2-m^2)[(k-p)^2-m^2]}\Bigg]	\nonumber\\
	\displaybreak[0]
& - 2d^\kappa_{\,\,\,\lambda}q^\alpha\frac{\partial}{\partial k^\alpha}\Bigg[ \frac{k^\lambda k^\tau p^\sigma k_\zeta (\delta^\zeta_{\,\,\,\mu}\varepsilon_{\kappa\tau\nu\sigma}+ \delta^\zeta_{\,\,\,\nu}\varepsilon_{\kappa\tau\sigma\mu}+ \delta^\zeta_{\,\,\,\sigma}\varepsilon_{\kappa\tau\mu\nu})}{(k^2-m^2)^2[(k-p)^2-m^2]}	\nonumber\\
	\displaybreak[0]
&\qquad\qquad\qquad + \frac{(k-p)^\lambda k^\tau p^\sigma (k-p)_\zeta (\delta^\zeta_{\,\,\,\mu}\varepsilon_{\kappa\sigma\tau\nu}+ \delta^\zeta_{\,\,\,\nu}\varepsilon_{\kappa\sigma\mu\tau}- \delta^\zeta_{\,\,\,\tau}\varepsilon_{\kappa\sigma\mu\nu})}{(k^2-m^2)[(k-p)^2-m^2]^2} \Bigg]	\nonumber\\
	\displaybreak[0]
& - d^\kappa_{\,\,\,\lambda}\varepsilon_{\kappa\rho\nu\mu}q^\alpha\frac{\partial}{\partial k^\alpha}\Bigg[ \frac{k^\lambda k^2 (k+p)^\rho}{(k^2-m^2)^2[(k-p)^2-m^2]} - \frac{(k-p)^\lambda (k^2-p^2) k^\rho}{(k^2-m^2)[(k-p)^2-m^2]^2} \Bigg]	\nonumber\\
	\displaybreak[0]
& + m^2 d^\kappa_{\,\,\,\lambda}\varepsilon_{\sigma\kappa\mu\nu}q^\alpha\frac{\partial}{\partial k^\alpha}\Bigg[ \frac{k^\lambda(k+p)^\sigma}{(k^2-m^2)^2[(k-p)^2-m^2]} + \frac{(k-p)^\lambda (k-2p)^\sigma}{(k^2-m^2)[(k-p)^2-m^2]^2} \Bigg] \nonumber\\
	\displaybreak[0]
& + (\mu;p\leftrightarrow\nu;q) \Bigg\}.
\end{align}
\end{widetext}
%
Making use of Gauss' theorem to transform the integrals (\ref{before-gauss}) into surface integrals%
\footnote{For this purpose, we Wick rotate to Euclidean space, $k_o=ik_4$ and $d^4k = id^4k_E$, such that $\int_V d^4k_E \frac{\partial}{\partial k^\alpha}\big(\dots\big)^{\alpha} = \oint_\mathcal{S} d\Omega\, k^2 k_\alpha \big(\dots\big)^{\alpha},$ where the last integration is along the solid angle of a 3-dimensional surface.}%
, and evaluating it at an isotropic surface at $k\rightarrow\infty$, reveals that the difference inside the third square brackets vanishes at the boundary and the surface integral of the terms inside the fourth square brackets goes to zero as $k\rightarrow\infty$. Expression (\ref{before-gauss}) simplifies to:
\begin{align}\label{after-gauss}
(p+q)^\rho & \Gamma_{\rho\mu\nu}(p,q)	\nonumber\\
& = - 4e^3d^\lambda_{\,\,\,\nu}\varepsilon_{\lambda\sigma\mu\rho}(p^\sigma q^\alpha - p^\alpha q^\sigma)\oint_{\cal S} \frac{d\Omega}{(2\pi)^4} \frac{k_\alpha k^\rho}{k^2}	\nonumber\\
	\displaybreak[0]
&\quad + 8e^3d^\kappa_{\,\,\,\lambda}\big(2\delta^\zeta_{\,\,\,\mu}\varepsilon_{\kappa\tau\nu\sigma} + 2\delta^\zeta_{\,\,\,\nu}\varepsilon_{\kappa\tau\sigma\mu} + \delta^\zeta_{\,\,\,\sigma}\varepsilon_{\kappa\tau\mu\nu}	\nonumber\\
	\displaybreak[0]
&\quad\quad\quad\quad\quad - \delta^\zeta_{\,\,\,\tau}\varepsilon_{\kappa\sigma\mu\nu}\big)p^\sigma q^\alpha \oint_{\cal S} \frac{d\Omega}{(2\pi)^4} \frac{k_\alpha k^\tau k^\lambda k_\zeta}{k^4}	\nonumber\\
&\quad + (\mu;p\leftrightarrow\nu;q).
\end{align}
The surface integrals are easily evaluated at isotropic momenta $k_\mu\rightarrow\infty$ by arguments of Lorentz covariance. The result is
\begin{align}\label{divergence-zero-shift}
(p+q)^\rho\Gamma_{\rho\mu\nu}(p,q) = & -\frac{e^3}{12\pi^2}\left( d^\lambda_{\,\,\,\mu}\varepsilon_{\lambda\nu\rho\sigma} + d^\lambda_{\,\,\,\nu}\varepsilon_{\mu\lambda\rho\sigma}\right.	\nonumber\\
	\displaybreak[0]
& \left.- d^\lambda_{\,\,\,\rho}\varepsilon_{\mu\nu\lambda\sigma} - d^\lambda_{\,\,\,\sigma}\varepsilon_{\mu\nu\rho\lambda}  \right)p^\rho q^\sigma	\nonumber\\
	\displaybreak[0]
& + \frac{e^3}{12\pi^2} d^{\kappa\lambda} \left( \varepsilon_{\kappa\lambda\nu\rho}\eta_{\mu\sigma} - \varepsilon_{\kappa\lambda\nu\sigma}\eta_{\mu\rho}\right.	\nonumber\\
	\displaybreak[0]
& \left. + \varepsilon_{\kappa\lambda\rho\mu}\eta_{\nu\sigma} - \varepsilon_{\kappa\lambda\sigma\mu}\eta_{\nu\rho} \right)p^\rho q^\sigma.
\end{align}

Expression (\ref{divergence-zero-shift}) is not the whole story because there is an internal momenta routing ambiguity in the triangular diagram, Fig. \ref{fig:triangular-two}. We may relabel all internal momenta by adding a constant vector $a$, \textit{i.e.},
\begin{equation}
k \rightarrow k+a,	\quad \textrm{where}	\quad a = \alpha q + (\alpha - \beta)p,
\end{equation}
with $\alpha$ and $\beta$ arbitrary constants, and ask what would change in our results. If the integral describing this process was perfectly finite, nothing would change, but it is actually linearly divergent and a relevant surface term emerges as a result of this routing ambiguity. Therefore, we should substitute result (\ref{divergence-zero-shift}) by another one which takes this ambiguity into account, \textit{i.e.}, one valid for an arbitrary shift of internal momenta, $\Gamma_{\rho\mu\nu}(p,q;a)$. A practical way of finding $\Gamma_{\rho\mu\nu}(p,q;a)$ is to first compute the surface term $\Delta_{\rho\mu\nu}(a)$,
\begin{equation}\label{surface}
\Delta_{\rho\mu\nu}(a) = \Gamma_{\rho\mu\nu}(p,q;a) - \Gamma_{\rho\mu\nu}(p,q;0),
\end{equation}
and use it to obtain
\begin{equation}\label{div-arb-shift}
(p+q)^\rho\Gamma_{\rho\mu\nu}(p,q;a) = (p+q)^\rho\left[ \Gamma_{\rho\mu\nu}(p,q;0) + \Delta_{\rho\mu\nu}(a) \right],
\end{equation}
where the first term at the right-hand side is already known --- it is Eq. (\ref{divergence-zero-shift}) ---, and the second one is just the divergence of (\ref{surface}). After that, we will have a result with explicit dependence on the routing ambiguity. Since our physical theory relies upon gauge invariance, we fix the ambiguity by requiring this symmetry to hold.

\begin{widetext}
Computation of $\Delta_{\rho\mu\nu}(a)$ follows very similarly the derivation of (\ref{divergence-zero-shift}). Thus, generalization of Eq. (\ref{divergence-zero-shift}) to consider an arbitrary shift of internal momenta reads:
\begin{align}\label{div-arb-shift-result}
(p+q)^\rho \Gamma_{\rho\mu\nu}(p,q;a) = & -(1+\beta)\frac{e^3}{12\pi^2}\left( d^\lambda_{\,\,\,\mu}\varepsilon_{\lambda\nu\rho\sigma} + d^\lambda_{\,\,\,\nu}\varepsilon_{\mu\lambda\rho\sigma} - d^\lambda_{\,\,\,\rho}\varepsilon_{\mu\nu\lambda\sigma} - d^\lambda_{\,\,\,\sigma}\varepsilon_{\mu\nu\rho\lambda}  \right)p^\rho q^\sigma	\nonumber\\
	\displaybreak[0]
& + (1+\beta)\frac{e^3}{12\pi^2} d^{\kappa\lambda} \left( \varepsilon_{\kappa\lambda\nu\rho}\eta_{\mu\sigma} - \varepsilon_{\kappa\lambda\nu\sigma}\eta_{\mu\rho} + \varepsilon_{\kappa\lambda\rho\mu}\eta_{\nu\sigma} - \varepsilon_{\kappa\lambda\sigma\mu}\eta_{\nu\rho} \right)p^\rho q^\sigma	\nonumber\\
	\displaybreak[0]
& - \beta\frac{e^3}{6\pi^2}d^\lambda_{\,\,\,\rho}\varepsilon_{\kappa\lambda\mu\nu}(p^\rho p^\kappa - q^\rho q^\kappa)	\nonumber\\
	\displaybreak[0]
& - \beta\frac{e^3}{12\pi^2}d^{\kappa\lambda}\left[ \varepsilon_{\kappa\lambda\mu\nu}(p^2-q^2) + \varepsilon_{\kappa\lambda\nu\rho}(p_\mu p^\rho - q_\mu q^\rho) + \varepsilon_{\kappa\lambda\rho\mu}(p_\nu p^\rho - q_\nu q^\rho) \right],
\end{align}
where terms proportional to $\beta$ are contributions coming from $(p+q)^\rho \Delta_{\rho\mu\nu}(a)$. This is the left-hand side of expression (\ref{triang-first-eq}), but it is not very illuminating in this form, so we recast it in terms of the vertex functional and fields:
\begin{align}\label{div-arb-shift-result-fields}
w_{\rm g}(x) \Gamma = & -(1+\beta)\frac{e^3}{24\pi^2}i\left( d^\lambda_{\,\,\,\mu}\varepsilon_{\lambda\nu\rho\sigma} - d^\lambda_{\,\,\,\nu}\varepsilon_{\mu\lambda\rho\sigma} + d^\lambda_{\,\,\,\rho}\varepsilon_{\mu\nu\lambda\sigma} - d^\lambda_{\,\,\,\sigma}\varepsilon_{\mu\nu\rho\lambda}  \right)\partial^\mu A^\nu \partial^\rho A^\sigma	\nonumber\\
	\displaybreak[0]
& + (1+\beta)\frac{e^3}{24\pi^2}i d^{\kappa\lambda} \left( \varepsilon_{\kappa\lambda\mu\nu}\eta_{\rho\sigma} + \varepsilon_{\kappa\lambda\rho\sigma}\eta_{\mu\nu} - \varepsilon_{\kappa\lambda\mu\sigma}\eta_{\nu\rho} - \varepsilon_{\kappa\lambda\rho\nu}\eta_{\sigma\mu} \right)\partial^\mu A^\nu \partial^\rho A^\sigma	\nonumber\\
	\displaybreak[0]
& + \beta\frac{e^3}{6\pi^2}id^\lambda_{\,\,\,\rho}\varepsilon_{\sigma\lambda\mu\nu}A^\mu \partial^\rho\partial^\sigma A^\nu + \beta\frac{e^3}{12\pi^2}id^{\kappa\lambda}\varepsilon_{\kappa\lambda\mu\nu}\left( A^\mu\Box{A^\nu} - A^\mu\partial^\nu\partial_\alpha{A^\alpha} - A_\alpha\partial^\alpha\,\partial^\mu{A^\nu} \right).
\end{align}
Finally, this expression can be further rewritten as:
\begin{align}\label{div-arb-shift-redefinition}
& w_{\rm g}(x) \Bigg\{ \Gamma - (1+\beta)\frac{e^3}{24\pi^2}i \int{ d^4y \Bigg[ (d^\lambda_{\,\,\,\mu}\varepsilon_{\lambda\nu\rho\sigma} - d^\lambda_{\,\,\,\nu}\varepsilon_{\mu\lambda\rho\sigma} )A^\mu A^\nu \partial^\rho A^\sigma + d^{\kappa\lambda}\varepsilon_{\kappa\lambda\mu\nu}(A^\mu A_\alpha\partial^\alpha A^\nu - A^\mu A_\alpha\partial^\nu A^\alpha - A^2\partial^\mu A^\nu) \Bigg] }	\Bigg\}	\nonumber\\
& = (1+3\beta)\frac{e^3}{24\pi^2}i\Bigg[(d^\lambda_{\,\,\,\rho}\varepsilon_{\mu\nu\lambda\sigma}+ d^\lambda_{\,\,\,\mu}\varepsilon_{\rho\nu\lambda\sigma}) A^{\sigma}\partial^\rho\partial^{\mu}A^{\nu} + d^{\kappa\lambda}\varepsilon_{\kappa\lambda\mu\nu}(A^{\mu}\Box A^{\nu} - A^\mu\partial^\nu\partial^\alpha{A_\alpha} - A_\alpha\partial^\alpha\partial^{\mu}A^{\nu})\Bigg].
\end{align}
\end{widetext}
The expression inside curly brackets in the left-hand side of (\ref{div-arb-shift-redefinition}) can be understood as a redefinition of the vertex functional but now comprising noninvariant counterterms so the right-hand side of (\ref{div-arb-shift-redefinition}) must be zero in order to ensure gauge invariance at this order. This fixes the parameter $\beta=-1/3$. Therefore, in the extended QED (\ref{qedex}), the one-loop three-photon vertex does not contribute to a potential gauge anomaly, \textit{i.e.}, in Eqs. (\ref{ward-anomalies}) and (\ref{triang-first-eq}) we have:
\begin{equation}\label{no-triang-anom}
\lambda^{(7)}_{\mu\nu\rho\sigma} = 0	\qquad	(\textrm{at one-loop order}).
\end{equation}

Concerning this result, at first order in Lorentz violation, due to the contraction with $F^{\mu\nu}F^{\rho\sigma}$, it was expected that $\lambda^{(7)}_{\mu\alpha\kappa\beta}$ could be made of only $(k_{AF})_{\mu\nu\rho\sigma}$ and $\varepsilon_{\mu\nu\rho\sigma}$. Clearly, there could be no $(k_{AF})_{\mu\nu\rho\sigma}$ present because the relevant diagrams only involve fermionic internal lines. In turn, no contribution involving only $\varepsilon_{\mu\nu\rho\sigma}$ arose because there was no ABBJ anomaly, which would manifest itself from the trace part of $d^{\mu\nu}$ --- explicit evaluation (using dimensional regularization) actually reveals that the overall contribution of the trace part of $d^{\mu\nu}$ vanishes for $\Gamma_{\rho\mu\nu}(p,q)$ even before contraction with $(p+q)^\rho$. On dimensional grounds, $\lambda_{\mu\nu\rho\sigma}$ could only receive contribution from coefficients for Lorentz violation with mass dimension zero --- therefore coefficients of $M_1$ are expected to give no contribution to the anomaly (\ref{triang-first-eq}) at any order in Lorentz violation ---, and, on symmetry grounds, from $C$- and $PT$-even ones. At first order in Lorentz violation, the only fermionic coefficient appearing in (\ref{qedex}) with this properties is $c_{\mu\nu}$, but it does not contribute to the anomaly because its related $C$-symmetry vanishes the sum depicted in Fig. \ref{fig:triangular-two}. 

We are in position to discuss multiloop contributions along with nonlinear ones from coefficients for Lorentz violation. The standard ABBJ anomaly vanishes at one-loop and will not receive further higher order corrections due to the nonrenormalization theorem of Adler and Bardeen \cite{adler-bardeen-theorem}. Explicitly checked to first order above, we may also conjecture that no coefficient for Lorentz violation will contribute to the anomaly (\ref{triang-first-eq}) because the anomaly coefficient $\lambda^{(7)}_{\mu\nu\rho\sigma}$ is comprised of an overall $C$-even combination of Lorentz-violating coefficients and the generalized Furry's theorem guarantees that only $C$-odd photon three-point functions give nonvanishing contribution to the process, which may not contribute to the $C$-even anomaly.

At last, we should point out that in Ref. \cite{coleman-glashow} it was verified that addition of an isotropic Lorentz-violating coefficient $\epsilon$, leading to a Lagrangian for Weyl spinor $u$ coupled to gauge fields of the form $iu^\dagger\big[D_o-(1-\frac{1}{2}\epsilon)\vec{\sigma}\cdot\vec{D}\big]u$, results in the very same anomaly of ABBJ, and for abelian gauge fields the anomaly vanishes, to all orders due to the Adler-Bardeen theorem, as in the standard Lorentz invariant case. Comparison with our case reveals this is analogous to considering (\ref{qedex}) with only Lorentz-violating coefficient the isotropic part (\textit{i.e.}, the trace part) of $d_{\mu\nu}$, and our results agree with \cite{coleman-glashow}. The nonisotropic contribution from $d_{\mu\nu}$ leads to an anomaly similiar to the ABBJ but with Lorentz-violating anomaly coefficients, as can be seen in (\ref{div-arb-shift-redefinition}), and we checked it vanishes to one-loop order. It would be interesting to study how the Adler-Bardeen theorem generalizes to this case, but this is beyond the scope of this work. As suggested before, there is a good indication the anomaly indeed vanishes to all orders.

\section{Summary}
\label{section:conclusions}

In this work we dealt with the search for possible gauge anomalies in a Lorentz-violating QED extension. Continuing the analyses of Refs. \cite{kost-1-loop,algebraic-renorm}, we made further progress by explicitly verifying the gauge invariance of the three-photon vertex to one-loop order and by presenting a conjecture stating the absence of gauge anomalies coming from this process to all orders, relying on the argument that generalized Furry's theorem may prevent the anomaly coefficient to receive nonvanishing contributions from coefficients for Lorentz violation. Explicit verifications, or a deeper analysis, of this conjecture would be of great interest along with an all-orders proof of the vanishing of the others anomaly coefficients, paving the way for a verification of renormalizability of the model to all orders in perturbation theory.

\section*{Acknowledgements}
The authors would like to thank Prof. V.A. Kosteleck\'{y} for very illuminating discussions and wise advices, O.M. Del Cima and O. Piguet for useful discussions in the preliminary stages of this work, and an anonymous referee for pointing out Ref. \cite{coleman-glashow} and its relation to our work. Financial support for this work was provided by CAPES.



\end{document}